\documentclass[preprint,superscriptaddress,showpacs,preprintnumbers,amsmath,amssymb,aps,pre]{revtex4}
\usepackage{graphicx}
\usepackage{color}
\begin{document}
\title{Macroscopic crack propagation in brittle heterogeneous materials analyzed in Natural Time}
\author{N. V. Sarlis}
\affiliation{Solid State Section and Solid Earth Physics
Institute, Physics Department, University of Athens,
Panepistimiopolis, Zografos 157 84, Athens, Greece}
\author{E. S. Skordas}
\affiliation{Solid State Section and Solid Earth Physics
Institute, Physics Department, University of Athens,
Panepistimiopolis, Zografos 157 84, Athens, Greece}
\author{P. A. Varotsos}
\thanks{{\bf Correspondence to:} P. Varotsos (pvaro@otenet.gr)}
\affiliation{Solid State Section and Solid Earth Physics
Institute, Physics Department, University of Athens,
Panepistimiopolis, Zografos 157 84, Athens, Greece}

\begin{abstract}
Here, we analyze in natural time $\chi$, the slow propagation of a
macroscopic crack in brittle heterogeneous materials through
sudden jumps and energy release events which are power law
distributed with universal exponents. This macroscopic crack
growth is currently believed to exhibit similar characteristics
with the seismicity associated with earthquakes. Considering that
the crack front is self affine and exhibits Family-Vicsek
universal scaling, we show that the variance $\kappa_1
(\equiv\langle \chi^2\rangle -\langle \chi \rangle^2)$ of natural
time is equal to 0.0686, which almost coincides with the value
$\kappa_1\approx 0.07$ obtained from the seismicity preceding
major earthquakes. This sheds light on the determination of the
occurrence time of an impending mainshock.
\end{abstract}


 \maketitle

\section{Introduction}
The effect of material heterogeneities on their failure
properties, which has been extensively studied\cite{ALA06}, still
remains one of the challenges among the unresolved questions in
fundamental physics\cite{BON08}. In a cluster issue on
Fracture\cite{BOU09}, several topics in the new avenues of
investigation concerning the basic mechanisms leading to
deformation and failure in heterogeneous materials, particularly
in non-metals, have been reviewed.

In that issue, Bonamy\cite{BON09} focused on the following two
aspects: First, the morphology of fracture surfaces (for an
earlier review see Ref. \cite{BOU97}), which constitutes  a
signature\cite{BON06} of the complex damage and fracture processes
occurring at the microstructure scale that lead to the failure of
a given heterogeneous material. Second, on the dynamics of cracks;
in particular, in heterogeneous materials, under slow external
loading, the crack propagation displays a jerky dynamics with
sudden jumps spanning over a broad range of length
scales\cite{MAL01,SAN04,MAR06}. Such complex dynamics, also called
`crackling noise'\cite{SET01}, is suggested from the acoustic
emission that accompanies the failure of various materials (e.g.
see Ref. \cite{GAR97,DAV07}) and, at a much larger scale, the
seismic activity associated with earthquakes (see also below).

Several experiments and field observations revealed that brittle
failure in materials exhibits scale-invariant features. In
particular: (a)roughness of cracks exhibits self-affine
morphological features, characterized by roughness exponents,
(b)the energy distribution of the discrete events observed in
crackling dynamics form a power laws with no characteristic scale.
Some of these scale-invariant morphological features and
scale-free energy distributions are universal. Others are not.
After reviewing the available experiments as well as the lattice
simulations,  it was concluded\cite{BON09} that the following two
cases should be distinguished:

(A)Damage spreading processes within a brittle material that {\em
precede} the initiation of a macroscopic crack. These
micro-fracturing events release energy impulses,  during this
transient damage spreading, that are power law distributed, but
the associated exponents are {\em non-universal}.

(B)Macroscopic crack growth within a brittle material. When this
propagation is slow enough, it exhibits an intermittent crackling
dynamics, with sudden jumps and energy release events the
distributions of which forms power law with apparent {\em
universal} exponents. Crack growth leads to rough fracture
surfaces that display Family-Vicsek universal scaling\cite{FAM91}
far enough from crack initiation.

It has been demonstrated\cite{BON08,BON09} that the crackling
dynamics in the observations performed in case (B) (for example,
the steady slow crack growth along a weak heterogeneous plane
within a transparent plexiglass block\cite{MAL06}) seems to be
captured quantitatively through a {\em stochastic} description
derived from linear elastic fracture mechanics (LEFM) extended to
disordered materials.  The role of temperature in this stochastic
LEFM description of crack growth has been also 
studied\cite{PON09B} and a creep law to relate crack velocity with
the stress intensity factor $K_1$  (see below) in the subcritical
failure regime has been proposed. This was found to describe
rather well experiments of paper peeling\cite{KOI07} and
subcritical crack growth in sandstone.\cite{PON09B}

The necessity of using a stochastic description could be
summarized, in simple words, as follows: According to Griffith's
theory\cite{GRI21}, assuming that the mechanical energy $G$
released as a fracture occurs is entirely dissipated within a
small zone at the crack tip and defining the fracture energy
$\Gamma$ as the energy needed  to create two crack surfaces of a
unit area, under the quasistatic condition, the local crack
velocity $v$ is assumed to be proportional to the excess energy,
$G-\Gamma$, locally released: $v/\mu = G-\Gamma$ where $\mu$ is
the effective mobility of the crack front. At the onset of crack
propagation ($v=0$), we have $G=\Gamma$ and $G$ is interrelated
with the stress intensity factor\cite{GAO86} $K_1$, determining
the singular stress field at the crack tip, through $G=K_1^2/E$
where $E$ is the Young modulus of the material (while -in contrast
to the early suggestion by Zener\cite{WER49,ZEN51}-  the
energy for the migration as well as for the formation of point
defects in solids\cite{KOS75} is governed, instead of $E$, by the bulk
modulus\cite{VAR997,VAR82,VAR82B,VAR80K133,VAR78,VAR08438}). In a
homogeneous medium $\Gamma$ is constant and an initially straight
crack front will be translated without being deformed. On the
other hand, in a heterogeneous material, which is of our interest
here, defects induce fluctuations in the local $\Gamma$. These
fluctuations induce local distortions in the crack front which in
turn generate local perturbations in $G$\cite{SCH95}. The
resulting effective force $F$ in this case is not constant
anymore, but given by the difference between the {\em mean} front
position and the one that would have been observed within the
homogeneous case\cite{BON08}. In such a {\em stochastic}
description, the onset of crack growth can be interpreted as a
critical transition (dynamic phase transition) between a stable
phase where the crack remains pinned by the material
inhomogeneities and a moving phase where the mechanical energy
available at the crack tip is sufficient to make the front
propagate\cite{SCH95,RAM97}(see also pages 254 \& 272 of Ref.
\cite{NEWBOOK}). As the crack grows, its mechanical energy is
reduced, thus the crack gets pinned again. This retroaction
process keeps the crack growth (provided it is slow enough) close
to the depinning transition at each time, thus the system remains
near the critical point during the whole propagation, in a similar
fashion as in self-organized criticality originally forwarded by
Bak, Tang and  Wiesenfeld\cite{BAK87}.

Recently, it has been shown that novel dynamic features hidden
behind the time series of complex systems in diverse fields (e.g.,
earth
sciences\cite{NAT02,NAT03A,NAT04,NAT05A,NAT05B,NAT05C,NAT06A,NAT08,SAR09B,SAR08,NAT09},
biology\cite{NAT02}, electrocardiograms\cite{NAT07,SAR09A},
physics\cite{SAR06}) can emerge if we analyze them in terms of a
newly introduced time domain, termed natural time
$\chi$\cite{NAT02}. This time domain, when employing the Wigner
function\cite{WIG32} and the generalized entropic measure proposed
by Tsallis\cite{TSA88}, it has been demonstrated\cite{ABE05} to be
 optimal for enhancing the signal's
localization in the time frequency space\cite{COH94}, which
conforms to the desire to reduce uncertainty and extract signal
information as much as possible. Natural time analysis enables the
study of the dynamic evolution of a complex system and identifies
when the system approaches the critical point. This occurs when
the value of the variance of natural time $\kappa_1 (\equiv
\langle \chi^2 \rangle -\langle \chi \rangle ^2)$ (see Section 2)
becomes
equal\cite{NAT02,NAT03A,NAT04,NAT05A,NAT05B,NAT05C,NAT06A,NAT08,SAR09B,SAR08}
to 0.070.

In view of the aforementioned analogy between the onset of crack
propagation and the critical dynamic transition, we investigate
here  for the first time the natural time analysis of macroscopic
crack growth within a disordered brittle material. This
investigation is of key importance, if we consider the following
two independent recent findings: First, as also pointed out by
Bonamy\cite{BON09}, quite surprisingly, seismicity associated with
earthquakes seems to belong to case (B) mentioned above and
exhibits quantitatively the same statistical scaling features as
the observed in laboratory experiments of interfacial crack growth
along weak disordered interfaces\cite{GRO09}. Second, it has been
empirically found  for major earthquakes in
Greece\cite{NAT02,NAT05C,NAT06A,NAT08,SAR09B} and
Japan\cite{UYE09}, including a sequence of major earthquakes which
occurred in Greece during 2008\cite{SAR08,UYE08,UYE10}, that the
occurrence time of a main shock can be determined by analyzing in
natural time the seismicity (for various magnitude thresholds, see
Appendix) that occurs after the detection of precursory electric
signals, termed Seismic Electric Signals (SES) activities, which
exhibit infinitely ranged temporal correlations ({\em critical
dynamics})\cite{NAT02,NAT03A,NAT04,NAT05A,NAT09}.
 SES  are transient low frequency
($\leq$ 1Hz) signals preceding
earthquakes\cite{VAR96B,NAT02A,TECTO13,VAR93B,PNAS} since they are
emitted when the stress in the focal region reaches a {\em
critical} value before the failure\cite{VARBOOK,VAR98}. These
signals, for earthquakes with magnitude 6.5 or larger, are
accompanied by detectable magnetic field
variations\cite{VAR01,VAR03}.

The paper is structured as follows: Section 2 provides a brief
description of natural time. Section 3 presents the analysis of
the macroscopic crack propagation in natural time; the stability
of the results of which is studied in Section 4. A brief
discussion follows in Section 5, while the main conclusions are
summarized in Section 6. Finally, an Appendix is provided which
explains how the present results can be applied to real seismic
data that precede the occurrence of major earthquakes.

\section{Natural time background}
In a time series comprising $N$ events, the {\em natural time}
$\chi_k = k/N$ serves as an index\cite{NAT02} for the occurrence
of the $k$-th event. The evolution of the pair ($\chi_k, Q_k$) is
studied, where $Q_k$ denotes a quantity proportional to the energy
released in the $k$-th event. For example, for dichotomous
signals, which is frequently the case of SES
activities\cite{NAT02,NAT03A,NAT04,NAT05A}, $Q_k$ stands for the
duration of the $k$-th pulse. In  the analysis of
seismicity\cite{NAT02,NAT05C,NAT06A,NAT08,SAR09B,SAR08}, $Q_k$ may
be considered as the seismic moment $M_{0 k}$ of the $k$-th event,
since  $M_{0}$ is roughly proportional to the energy released
during an earthquake\cite{KAN79}. The normalized power spectrum
$\Pi(\omega )\equiv | \Phi (\omega ) |^2 $ was
introduced\cite{NAT02} where
\begin{equation}
\label{eq3} \Phi (\omega)=\sum_{k=1}^{N} p_k \exp \left( i \omega
\frac{k}{N} \right)
\end{equation}
and $p_k=Q_{k}/\sum_{n=1}^{N}Q_{n}$, $\omega =2 \pi \phi$; $\phi$
stands for the {\it natural frequency}. The continuous function
$\Phi (\omega )$ should {\em not} be confused with the usual
discrete Fourier transform, which considers only its values at
$\phi=0,1,2,\ldots$. In natural time analysis\cite{NAT02},
 the properties of
$\Pi(\omega)$ or $\Pi(\phi)$ are studied  for natural frequencies
$\phi$
 less than 0.5, since in
this range of $\phi$, $\Phi(\omega)$  reduces
 to a {\em characteristic function} for the
probability distribution $p_k$  in the context of probability
theory: for $\omega \rightarrow 0$, all the moments of the
 distribution of $p_k$ can be estimated from $\Phi (\omega )$ (see
p.499 of Ref. \cite{FEL71}).  Equation(\ref{eq3}) in this limit
leads to $\Pi (\omega )\approx 1-\kappa_1 \omega^2$ which reflects
that the quantity $\kappa_1$ equals the variance of $\chi$:
\begin{equation}\label{eqk1}
\kappa_1=\langle \chi^2 \rangle -\langle \chi \rangle^2,
\end{equation}
 where $\langle f( \chi) \rangle = \sum_{k=1}^N p_k
f(\chi_k )$. This, of course, may vary upon the occurrence of each
new event.

 When the system enters the
{\em critical} stage, the following relation
holds\cite{NAT02,NEWBOOK}
\begin{equation}
\Pi ( \omega ) = \frac{18}{5 \omega^2} -\frac{6 \cos \omega}{5
\omega^2} -\frac{12 \sin \omega}{5 \omega^3}. \label{fasma}
\end{equation}
For $\omega \rightarrow 0$, Eq.(\ref{fasma}) leads
to\cite{NAT02,NEWBOOK} $\Pi (\omega )\approx 1-0.07 \omega^2$
which reflects that the variance of $\chi$ is given by
\begin{equation}
 \kappa_1=\langle \chi^2 \rangle -\langle \chi \rangle
^2=0.07.
\end{equation}

\section{Macroscopic crack propagation in natural time}

In the frame\cite{BON09} of the stochastic LEFM description of
macroscopic crack growth in heterogeneous materials (case (B)
discussed in Section 1), the crack front $f(z,t)$ (e.g., Fig.14 of
Ref. \cite{BON09}) is self-affine and exhibits Family-Vicsek
dynamic scaling up to a correlation length $\xi \propto \Delta
t^{1/\kappa}$, i.e.,
\begin{equation}\label{eq0}
\langle \left[ f(z+\Delta z, t+\Delta t)- f(x,t)
\right]\rangle^{1/2} \propto \Delta t^{\zeta_H/\kappa} g
\left(\frac{\Delta z}{\Delta t^{1/\kappa}}\right),
\end{equation}
where $g(u)$ is a scaling function and $\zeta_H$ and $\kappa$
refer to the roughness exponent and the dynamic exponent,
respectively. As the front propagation occurs through avalanches
between two successive pinned configurations (see Fig.16(a) of
Ref. \cite{BON09}), an
 avalanche of size $S$ results in an increment $\Delta \langle f \rangle=S/L$ (where $L$
is the system size) in the mean crack length $\langle f \rangle$.
The mechanical energy $E$ released during the avalanche is
also\cite{BON09} proportional to $S$ (see Eq.(20) and Fig.16(c) of
Ref. \cite{BON09}). Assuming that the increments $\Delta \langle f
\rangle$ are proportional to the standard deviation $\langle
\left[ f(z+\Delta z, t+\Delta t)- f(x,t) \right]\rangle^{1/2}$, we
obtain from Eq.(\ref{eq0}) that
\begin{equation}\label{eq1}
E\propto S \propto L \Delta \langle f \rangle \propto \langle
\left[ f(z+\Delta z, t+\Delta t)- f(x,t) \right]\rangle^{1/2}
\propto \Delta t^{\zeta_H/\kappa}.
\end{equation}
The most recent evaluations of $\zeta_H$ and
$\kappa$\cite{ROS02,DUE07} result in\cite{BON09}
$\zeta_H=0.385(5)$ and $\kappa=0.770(5)$, giving rise to a ratio
$\lambda=\zeta_H/\kappa=0.500(7)$, i.e., a square root growth law
for the energy emitted during the crack propagation. Equation
(\ref{eq1}) implies that in natural time $\chi$, the average
energy of the $k$-th avalanche $\langle E_k \rangle$ scales as
\begin{equation}
\langle E_k \rangle \propto k^\lambda. \label{ek}
\end{equation}
 Thus, we have $p_k \left(
=Q_k/\sum_{n=1}^{N}Q_{n} \right) \propto k^\lambda$, since $Q_k$
should be proportional to $E_k$, leading to $p(\chi ) \propto
\chi^\lambda$, i.e.,
\begin{equation}\label{dyn}
 p(\chi)=(\lambda+1)\chi^\lambda
\end{equation}
 so that $\int_0^1 p(\chi) d\chi=1$.
Upon using Eq.(\ref{dyn}) for the estimation of the variance of
natural time, $\kappa_1=\int_0^1 \chi^2 p(\chi) d\chi -\left[
\int_0^1 \chi p(\chi) d\chi\right] ^2$, we find
\begin{equation} \label{eqk1dyn}
\kappa_1=\frac{\lambda+1}{\lambda+3}-\left(
\frac{\lambda+1}{\lambda+2} \right)^2
\end{equation}
Substituting the aforementioned value  $\lambda=0.500(7)$ in
Eq.(\ref{eqk1dyn}), we obtain $\kappa_1=0.0686(3)$. This almost
coincides with the value $\kappa_1 \approx 0.07$ empirically
found\cite{NAT02,NAT05C,NAT06A,NAT08,SAR09B,SAR08,UYE09} (see also
the Appendix) from the natural time analysis of the seismicity
before the occurrence of large earthquakes.

\section{The Stability of the result obtained in the previous section}

 The analysis of the
macroscopic crack propagation in natural time presented in the
previous Section, was made by using Eq.(\ref{ek}), which  suggests
that the expectation value $\langle E_k \rangle$ of the energy
$E_k$ should scale as
\begin{equation} \label{eq2}
\langle E_k \rangle \propto \sqrt{k}.
\end{equation}
In order to verify that the statistical character of
Eqs.(\ref{eq0}) and (\ref{eq1}) does not affect the validity of
the relation $\kappa_1 =0.0686$ for propagating macroscopic
cracks, we proceeded to Monte Carlo simulations assuming
\begin{equation}\label{model}
Q_k=E_k=r_k \sqrt{k} ,
\end{equation}
 where $r_k$ are independent and identically distributed variables such that $\langle E_k \rangle=\sqrt{k}$.
For each Monte Carlo calculation, $10^3$ realizations of the
stochastic process described by Eq.(\ref{model}) was performed.
For each realization, the resulting $E_k$ were analyzed in natural
time and the value of $\kappa_1$ as a function of $k$ was
determined (e.g., when $k=5$, only the first five $E_k,
k=1,2,\ldots,5$ were analyzed in natural time to obtain the
corresponding $\kappa_1$-value). Then, for each $k$, the mean
 and the standard deviation of the corresponding
$\kappa_1$-values were computed using the $10^3$ realizations.

%\begin{figure}
%\includegraphics{Figure1.eps}
%\caption{(color online) (a) Time-series of avalanche energies
%$E_k$ (thin line, right scale) obtained from Eq.(\ref{model})
%using exponentially distributed $E_k$ together with the
%corresponding evolution of $\kappa_1$ as a function of the number
%of avalanches $k$. (b)The average value (dark colors) and the $\pm
%\sigma$ intervals (light colors) of $\kappa_1$ as they result from
%Monte Carlo calculation for the case of exponentially (red),
%Poisson (blue) and uniformly (green) distributed $E_k$ (for
%details see Ref.\cite{EPAPS}).} \label{fig1}
%\end{figure}

\begin{figure}
\includegraphics{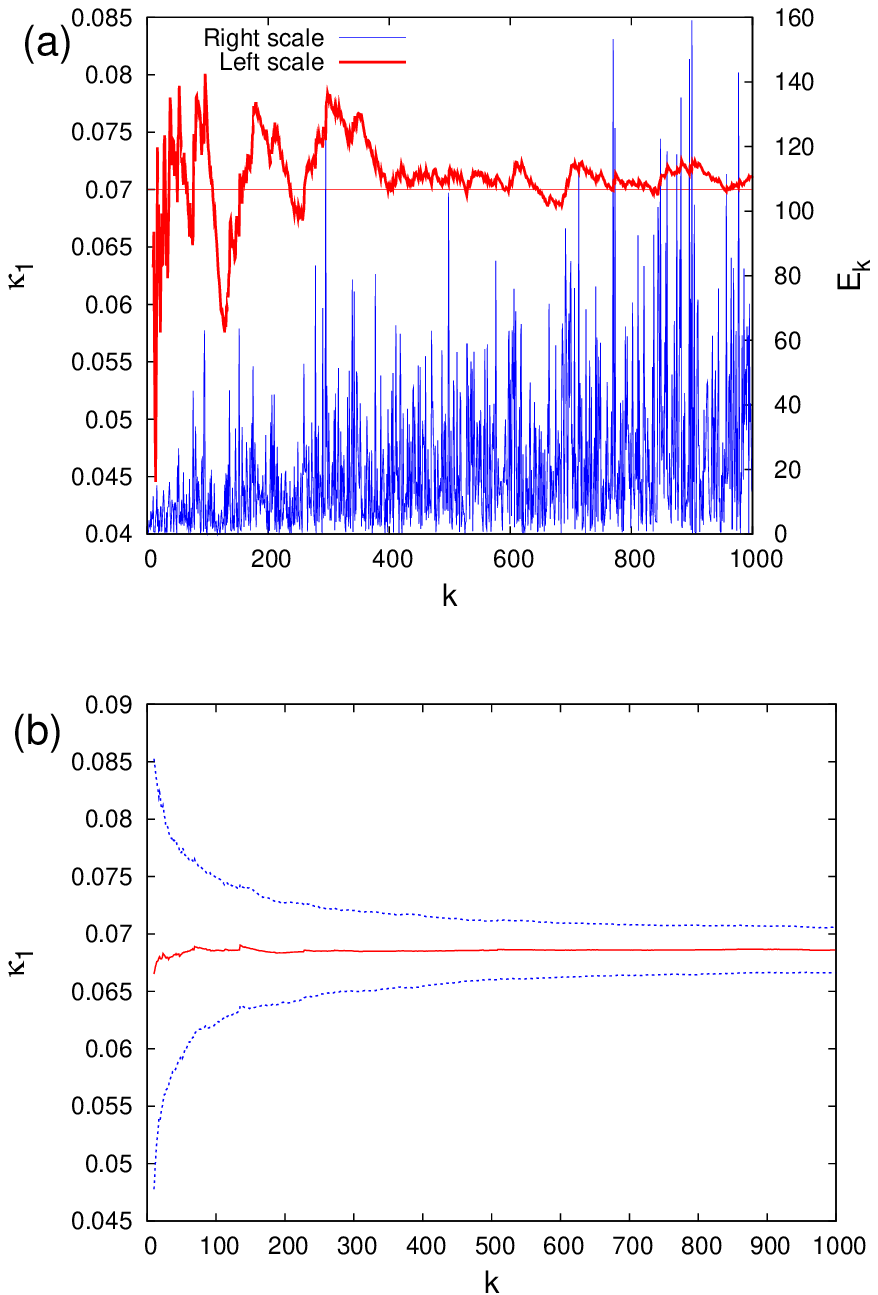}
\caption{(color online) (a) Time-series of avalanche energies
$E_k$ (thin blue line, right scale) obtained from Eq.(\ref{eq2})
using exponentially distributed $E_k$ together with the
corresponding evolution of $\kappa_1$ (thick red line, left scale)
as a function of the number of avalanches $k$. (b)The average
value (solid red) and the $\pm \sigma$ intervals (dotted blue) of
$\kappa_1$ as they result from Monte Carlo calculation of $10^3$
realizations of the processes shown in (a).} \label{f1}
\end{figure}

Figure \ref{f1}(a) shows with the thin blue line how an example
time-series of $E_k$, obtained from Eq.(\ref{model}) for $r_k$
exponentially distributed, is read in natural time. It exhibits an
intermittent behavior similar to that of Figs.2(a) and 16(c) of
Ref. \cite{BON09}, which depict the energy emission versus
conventional time from earthquakes and propagating cracks,
respectively. In the same figure, we also plot the corresponding
$\kappa_1$-value as a function of the order $k$ of the avalanche
(thick red solid line). Figure \ref{f1}(b) depicts the average
value of $\kappa_1$ (red solid line) together with the $\pm
\sigma$ (one standard deviation) intervals (blue dotted lines),
obtained from the Monte Carlo calculation of the process, an
example of which is shown in Fig.\ref{f1}(a). We observe that
these $\kappa_1$ values scatter around approximately 0.07, while
the average value saturates to 0.0686, as expected from the
analytical result of the previous Section.

\begin{figure}
\includegraphics{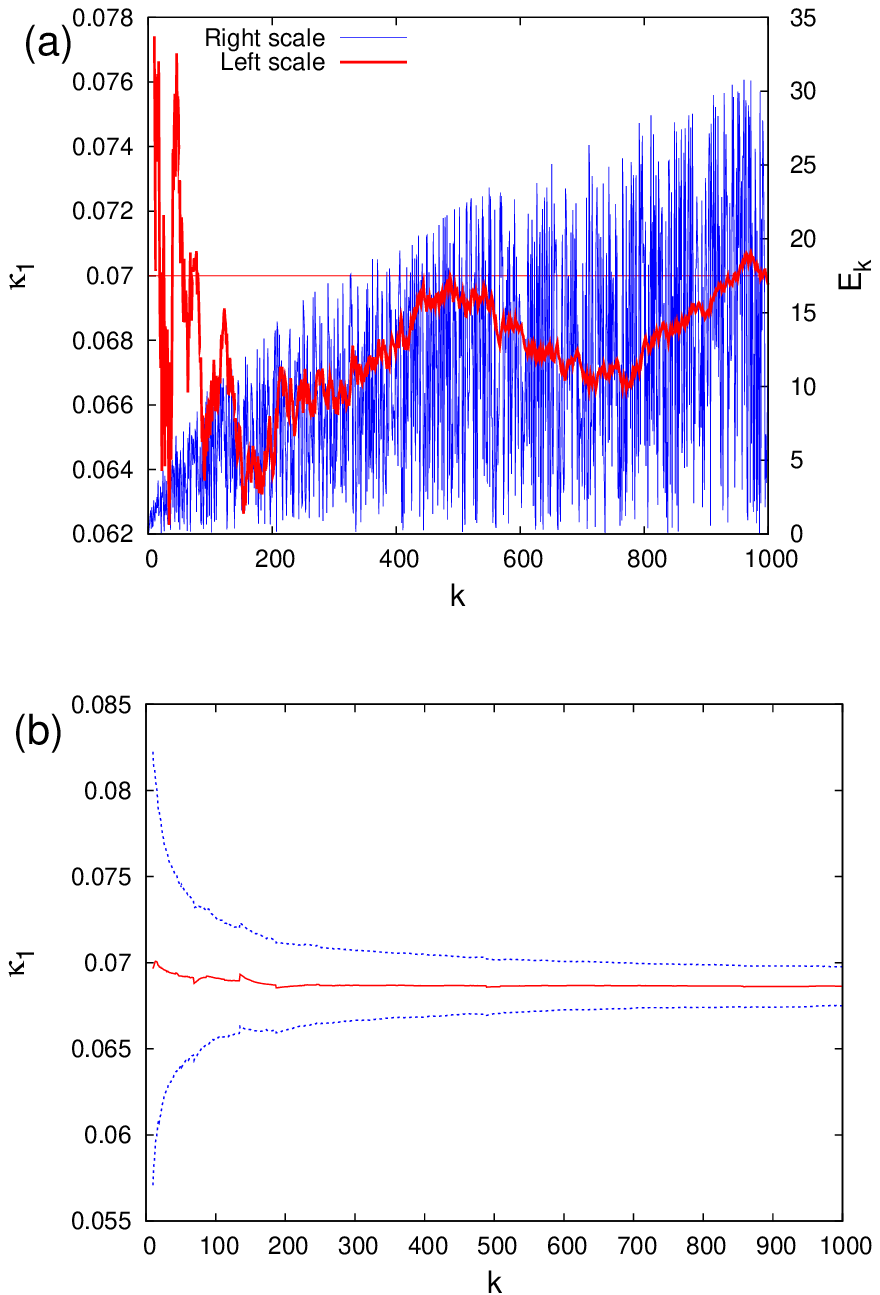}
\caption{(color online) (a) Time-series of avalanche energies
$E_k$ (thin blue line, right scale) obtained from Eq.(\ref{eq2})
using uniformly distributed $E_k$ together with the corresponding
evolution of $\kappa_1$  (thick red line, left scale) as a
function of the number of avalanches $k$. (b)The average value
(solid red) and the $\pm \sigma$ intervals (dotted blue) of
$\kappa_1$ as they result from Monte Carlo calculation of $10^3$
realizations of the processes shown in (a).} \label{f2}
\end{figure}

\begin{figure}
\includegraphics{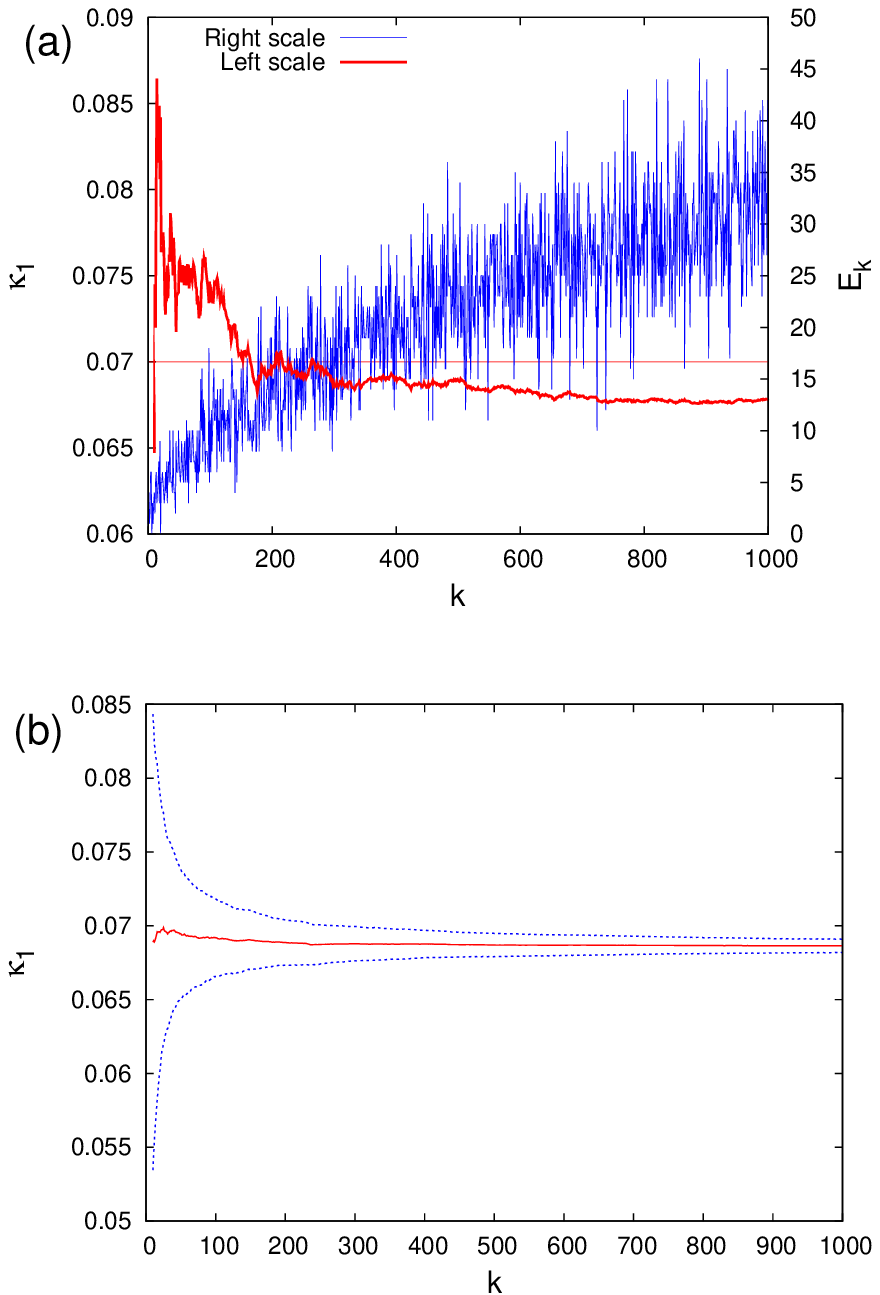}
\caption{(color online) (a) Time-series of avalanche energies
$E_k$ (thin blue line, right scale) obtained from Eq.(\ref{eq2})
using Poisson distributed $E_k$ together with the corresponding
evolution of $\kappa_1$ (thick red line, left scale) as a function
of the number of avalanches $k$. (b)The average value (solid red)
and the $\pm \sigma$ intervals (dotted blue) of $\kappa_1$ as they
result from Monte Carlo calculation of $10^3$ realizations of the
processes shown in (a).} \label{f3}
\end{figure}

Two additional monoparametric distributions of $E_k$ have been
also investigated: (a)uniformly distributed $E_k$ satisfying
Eq.(\ref{eq2}), see Fig.\ref{f2}, and (b)Poisson distributed
random variables $E_k$ obeying Eq.(\ref{eq2}), see Fig.\ref{f3}.
In addition, we note that when the quantities $r_k$, defined in
Eq.(\ref{model}), are Poisson distributed with a mean value
$\mu=\langle r_k \rangle$, a behavior intermediate between the
ones depicted in Figs.\ref{f1} and \ref{f3} -depending on whether
$\mu$ is small (with respect to unity) or large, respectively- is
found.

\begin{figure}
\includegraphics{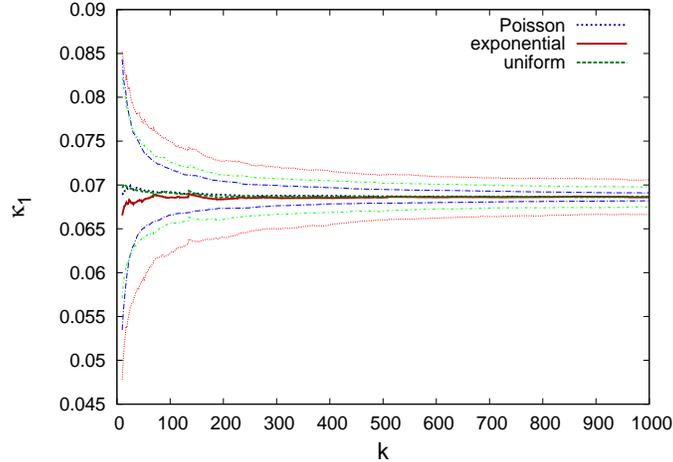}
\caption{(color online) The average value (dark colors) and the
$\pm \sigma$ intervals (light colors) of $\kappa_1$ as they result
from Monte Carlo calculation of $10^3$ realizations for
exponentially (red), uniformly (green) and Poisson (blue)
distributed $E_k$ that satisfy Eq.(\ref{eq2}).} \label{f4}
\end{figure}

The results of the aforementioned three monoparametric
distributions investigated, are summarized in Fig. \ref{f4}. In
all cases,  the mean value of $\kappa_1$ is around 0.070 and tends
to the value $\kappa_1=0.0686$ obtained analytically from
Eq.(\ref{eqk1dyn}). Among the three monoparametric distributions
studied, the exponential exhibits the highest variability while
Poisson the smallest. For example at $k=200$ the variability,
i.e., the ratio of the standard deviation over the mean value of
$\kappa_1$ is 6\%, 4\% and 2\% for the exponentially, uniformly
and Poisson distributed $E_k$, respectively. The latter is
understood from the following fact: as the average value $\langle
E_k \rangle \propto \sqrt{k}$ increases, the Poisson distribution
becomes almost Gaussian with standard deviation equal to
$\sqrt{\langle E_k \rangle}$ which is well localized resulting in
a small variability of $\kappa_1$ (see also Fig.\ref{f3}(a)).
Future research is needed to clarify which of these three
distribution is closer to the reality.

\section{DISCUSSION}

In view of the fluctuations seen in Figs.\ref{f1}(a), \ref{f2}(a),
and \ref{f3}(a), we now study
 the stability of $\kappa_1$ when  only one
realization of the process is available, as in the case of
seismicity (see Appendix). Such a stability is worthwhile to be
investigated since it may exist possibly stemming
 from the following two facts: (a)$\kappa_1$ exhibits\cite{NAT05B}
experimental (or Lesche\cite{LES82,LES04}) stability and (b)Equation
(\ref{eq2}) has a scale invariant feature.

Exponentially distributed $E_k$ satisfying Eq.(\ref{eq2}) exhibit,
as mentioned, the highest variability, e.g. 6\% at $k=200$ (see
Figs.\ref{f1} and \ref{f4}). Thus, hereafter, we adopt this
distribution for the purpose of our study. Focusing on the case of
seismicity, after a SES activity and before a mainshock, the
following question naturally arises: Since only one realization of
the process is known and in view of the variability of $\kappa_1$
(see Fig.\ref{f1}), could  it be possible to draw any conclusion
on the basis of $\kappa_1$? The following aspects may answer this
question: Let us consider a single realization of Eq.(\ref{eq2})
with exponentially distributed $E_k$, so that to obtain a series
$E_k=\epsilon_k, k=1,2, \ldots ,N$. We randomly select a large
number $M$ of subseries of $\epsilon_k$ of random length $\nu$,
which are analyzed in natural time to obtain $\kappa_1$. These
$\kappa_1$-values enable the construction of the  probability
density function (PDF) of $\kappa_1$. For example, let $\nu=3$,
and select the subseries $\epsilon_{k_1}$, $\epsilon_{k_2}$ and
$\epsilon_{k_3}$, according to their time order, i.e,
$k_1<k_2<k_3$. We can then find the corresponding $\kappa_1 \left[
=\sum_{k=1}^3 \chi_k^2 p_k -\left( \sum_{k=1}^3 \chi_k
p_k\right)^2\right]$ value by assuming only three events $k=1,2$
and 3, with $\chi_1=1/3$ and
$p_1=\epsilon_{k_1}/(\epsilon_{k_1}+\epsilon_{k_2}+\epsilon_{k_3})$,
$\chi_2=2/3$ and
$p_2=\epsilon_{k_2}/(\epsilon_{k_1}+\epsilon_{k_2}+\epsilon_{k_3})$
and $\chi_3=1$ and
$p_3=\epsilon_{k_3}/(\epsilon_{k_1}+\epsilon_{k_2}+\epsilon_{k_3})$.
Figure \ref{f5} depicts the PDFs of $\kappa_1$ obtained this way
from a single realization $\epsilon_k$ of $N=300$ events which are
exponentially distributed and satisfy Eq.(\ref{eq2}) for two
values of $M$ ($10^3$ or $10^4$) when $\nu$ is uniformly
distributed from 2 to 300. We observe that the resulting PDF of
$\kappa_1$ maximizes when $\kappa_1\approx 0.070$, i.e., close to
the $\kappa_1$-value obtained from Eq.(\ref{eqk1dyn})  for
$\lambda=0.5$. This result stems from the fact that the analysis
of the subseries maintains the {\em true}  time order of events
which constitutes  the {\em key point} of natural time. Attention
is drawn to the point  that if we destroy the order of the events
by {\em randomly} shuffling $\epsilon_k$ and then apply the same
procedure, the resulting PDF shown in black in Fig.\ref{f5} has
its peak displaced towards $\kappa_u=1/12\approx$0.0833. The
latter is the value of $\kappa_1$ corresponding to a ``uniform''
(u) distribution (as defined in Ref. \cite{NAT03A,NAT04,NAT05A},
e.g. when all $p_k$ are equal or $Q_k$ are positive independent
and identically distributed random variables of finite variance).

Thus, the aforementioned investigation showed that even when using
a single realization of the process described by Eq.(\ref{eq2})
with $E_k$ exponentially distributed, and select random subseries
of the process to be analyzed in natural time, the PDF deduced for
$\kappa_1$ maximizes at $\kappa_1 \approx 0.070$. This fact, after
recalling the point mentioned in Section 1 that seismicity
associated with earthquakes seems to belong\cite{BON09} to case
(B) (i.e., the macroscopic crack growth when this propagation is
slow enough),  sheds light on the usefulness of the natural time
analysis of the seismicity before a mainshock.   This analysis,
that is described in detail in the Appendix, considers the
time-series of seismicity above a magnitude threshold $M_{thres}$
that occurs  after the initiation of the SES activity (which
signifies that the system enters the critical
regime\cite{VARBOOK,VAR98}) in the area
candidate\cite{VAR91,DOL99,NEWBOOK,TEC05} to suffer a mainshock.
In addition, it considers the subseries corresponding to the
seismicity in all possible subareas of the candidate area, which
enables the  construction of the PDF of the resulting $\kappa_1$
values. It is found\cite{SAR08} that this PDF exhibits a maximum
at $\kappa_1\approx 0.070$ (in agreement with Fig.(\ref{f5})) when
approaching the time of occurrence of the main shock. In other
words, we set the natural time zero at the initiation time of the
SES activity, and then form time series of seismic events in
natural time (for every possible subarea), each time when a small
earthquake with $M \geq M_{thres}$ occurred, i.e., when the number
of events increases by one. Hence, Eq.(\ref{ek}) cannot be
misinterpreted as implying that the average energy of each event
is increasing {\em for ever} as a square root of the index $k$ of
the natural time, thus leading to an obviously unacceptable result
according to which we have to expect more and more powerful
earthquakes in the future. This equation is applied here only for
the small events that precede a main shock by starting (i.,e,
$k=0$) upon the initiation of the SES activity until just before
the main shock occurrence.  This behavior should not change upon
considering various magnitude thresholds since the process should
be scale invariant (see Appendix).

\section{CONCLUSIONS}

In summary, we studied in natural  time, the macroscopic crack
growth within a disordered brittle material when the propagation
is slow enough. It exhibits jerky dynamics with sudden jumps and
energy release events which are power law distributed  with
universal exponents. By considering that the crack front is
self-affine with Family-Vicsek dynamic scaling, we find that the
variance $\kappa_1(=\langle \chi^2 \rangle -\langle \chi
\rangle^2)$ is equal to $\kappa_1=0.0686(3)$. This, quite
interestingly, almost coincides with the value $\kappa_1 \approx
0.07$ obtained from the natural time analysis of the seismicity
that precedes major earthquakes.  This conforms with the current
aspect\cite{BON09}
 that the macroscopic crack growth exhibits similar features with the seismicity
 associated with earthquakes.  In other words, the present
result that $\kappa_1=0.0686$ for the slow propagation of a
macroscopic crack in heterogeneous materials through sudden jumps
(reminiscent of stick-slip phenomena), sheds light on the
following finding: After the detection of a SES activity (critical
dynamics), when following the dynamic evolution of the system by
computing $\kappa_1$ after each small earthquake -by means of the
natural time analysis of seismicity- and find that $\kappa_1
\approx 0.07$, we identify that the system approached the critical
point and the major mainshock occurs.

\appendix
\section{Application of the results obtained from the natural time
analysis of macroscopic crack propagation to real seismic data}
Here, we focus on the analysis of the seismicity after the
initiation of a SES activity and before a mainshock. Once a SES
activity has been recorded, which signifies that the system just
entered the critical regime\cite{VARBOOK,VAR98}, an estimation of
the area to suffer a mainshock can be obtained on the basis of the
so-called selectivity map\cite{VAR91,DOL99,NEWBOOK,TEC05} of the
station at which the SES observation was made. Thus, we have some
area, hereafter labelled A, in which we count the small events
(earthquakes) $e_i$ that occur after the initiation of the SES
activity. Each event $e_i$ is characterized by its location ${\bf
x}(e_i)$, the conventional time  of its occurrence $t(e_i)$, and
its magnitude ${\rm M}(e_i)$ or the equivalent seismic moment
$M_0(e_i)$ (e.g. see Ref. \cite{KAN79}). The index $i=1,2,\ldots,
N$ increases by one each time a new earthquake occurs within the
area A (cf. $\forall i, t(e_{i+1})> t(e_i)$). Thus, a set of
events $\cal{A}$ is formed until the mainshock occurs in A at
$i=N$.
 To be more precise, a family of
sets ${\cal A}_{{\rm M}_{thres}}$ of the earthquakes with
magnitude greater than or equal to $M_{thres}$ is formed, where
${\cal A}_{{\rm M}_{thres}}=\left\{ e_i \in {\cal A}: {\rm
M}(e_i)\geq {\rm M}_{thres}\right\}$ and the number of events in
${\cal A}_{{\rm M}_{thres}}$ is denoted by $|{\cal A}_{{\rm
M}_{thres}}|$. The set ${\cal A}_{{\rm M}_{thres}}$ becomes a
(time) ordered set by selecting the indices $j$ for its elements
$e_j$, $j=1,2,\ldots ,|{\cal A}_{{\rm M}_{thres}}|$  so that
$\forall j, t(e_{j+1})> t(e_j)$. Since earthquakes do not occur
everywhere within the area A but in some specific locations, we
may also define $R({\cal A}_{{\rm M}_{thres}})$ as the minimal
rectangular (in latitude and longitude) region in which the
epicenters of  the events of ${\cal A}_{{\rm M}_{thres}}$ are
located. Moreover, for a given ordered set ${\cal A}_{{\rm
M}_{thres}}$ the corresponding value of $\kappa_1({\cal A}_{{\rm
M}_{thres}})$  can be obtained by analyzing in natural time its
ordered elements $e_j$ ($\in {\cal A}_{{\rm M}_{thres}}$). This is
made by analyzing in natural time the pairs
$(\chi_j,Q_j)=(j/|{\cal A}_{{\rm M}_{thres}}|, M_0(e_j))$ where
$j=1,2,\ldots ,|{\cal A}_{{\rm M}_{thres}}|$.

\begin{figure}
\includegraphics{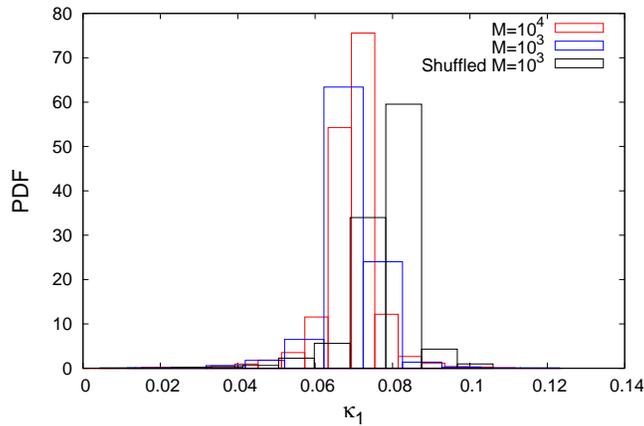}
\caption{(color online) The probability density functions (PDF) of
$\kappa_1$ obtained after randomly selecting $M=10^3$ (blue) or
$M=10^4$(red) subseries from a single realization of the process
described by Eq.(\ref{eq2}) using exponentially distributed $E_k$.
Both distributions are peaked close to 0.070. Once the events of
the original realization are shuffled randomly and then
 $M=10^3$ subseries are analyzed, the peak of the new PDF, shown in black, is displaced to the right.} \label{f5}
\end{figure}

The key point behind this approach, is the experimental (or
Lesche\cite{LES82,LES04})  stability
 that is satisfied\cite{NAT05B} by the variance $\kappa_1$  in combination with the conclusion drawn in
  Section 5.  We define a
{\em  proper } subset ${\cal P}_{{\rm M}_{thres}}$ of ${\cal
A}_{{\rm M}_{thres}}$ as a subset of ${\cal A}_{{\rm
M}_{thres}}\left( \supset {\cal P}_{{\rm M}_{thres}} \right)$ such
that it includes {\em all} the elements of ${\cal A}_{{\rm
M}_{thres}}$ that occurred after the SES initiation and before the
mainshock in $R({\cal P}_{{\rm M}_{thres}})$. Finally, we consider
the ensemble ${\cal E[ A}_{{\rm M}_{thres}}]$ of {\em all}
different  ${\cal P}_{{\rm M}_{thres}}$ which one can define from
a given  ${\cal A}_{{\rm M}_{thres}}$. For each member of this
ensemble, we can find the corresponding $\kappa_1$ value and, when
considering the totality of these values for the ensemble, we can
construct the PDF of $\kappa_1$. If the earthquakes under
consideration belong to a self-similar preparation process, we
expect that at the later stages of this process, where
Eq.(\ref{eq2}) might be valid, the PDF should exhibit a maximum at
$\kappa_1=0.070$. All the recent experimental results associated
with the major earthquakes in Greece (see Ref. \cite{SAR08})
indicate that  before the occurrence of a mainshock this PDF (or
equivalently the probability Prob($\kappa_1$) versus $\kappa_1$)
exhibits a maximum close to $\kappa_1=0.070$,  in agreement with
Fig.\ref{f5}.

\begin{figure}
\includegraphics{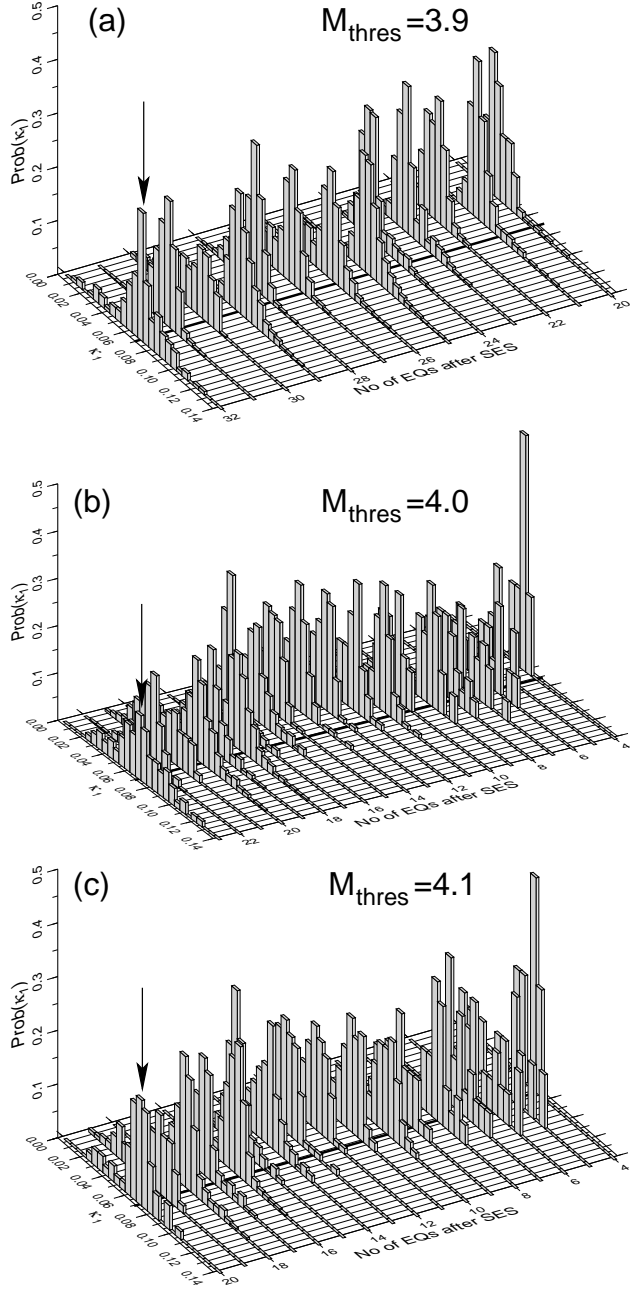}
\caption{(a):Prob($\kappa_1$) versus $\kappa_1$ of the
seismicity(subsequent to the long duration SES activity recorded
at Pirgos station during February 29 to March 2, 2008), for
$M_{thres}=3.9$,  within the SES selectivity map of Pirgos station
(which is the shaded area shown in Fig.8 of Ref.
\cite{sarlis-2008}). The arrow marks the maximum of
Prob($\kappa_1$) at $\kappa_1=0.07$ that occurred on 27 May  2008
(i.e., upon the occurrence of the 32nd event after the SES). This
maximum has been followed by the $M_w$6.4 mainshock on 8 June
2008.  (b) and (c) are the same as (a), but for $M_{thres}=4.0$
and $M_{thres}=4.1$, respectively. The arrows in (b) and (c)
indicate the maximum of Prob($\kappa_1$) at $\kappa_1\approx
0.070$ that also occurred on 27 May 2008.} \label{figa1}
\end{figure}

 As an
example, we summarize here the results obtained for the case of
the $M_w$6.4 earthquake that occurred at 38.0$^o$N21.5$^o$E at
12:25UT on  8 June  2008, which is the latest $M_w>6.0$ earthquake
in Greece. Figure 7(c) of Ref. \cite{sarlis-2008}(or Fig.3(b) of
Ref. \cite{SAR08}) depicts a long duration SES activity that was
recorded at Pirgos (Greece) measuring station from 29 February
2008 to 2 March 2008. This SES activity, as shown in Ref.
\cite{NAT09}, exhibits self-similar structure over five orders of
magnitude with a (self-similarity) exponent close to unity. By
studying the seismicity, after the initiation of this SES activity
in the grey shaded  area of Fig.8 of Ref. \cite{sarlis-2008} (or
Ref. \cite{SAR08}), which constitutes the SES selectivity map of
Pirgos station,
  we obtained (just after the occurrence of an
earthquake of magnitude 5.1 at 23:26UT on 27 May 2008) the
probabilities Prob($\kappa_1$) shown in Figs.\ref{figa1}(a), (b)
and (c) for $M_{thres}=3.9$, 4.0 and 4.1, respectively. These
distributions exhibit a maximum at $\kappa_1=0.070$ {\em
independent} of the magnitude threshold (as intuitively expected
for a self-similar preparation process). This fact was
reported\cite{sarlis-2008} on 29 May 2008, and eleven days later
the aforementioned $M_w$6.4 mainshock occurred within the area
(selectivity map of Pirgos station) specified in advance (cf. such predictions are issued only when the expected magnitude of the impending mainshock is 6.0 or larger, e.g., see Ref.\cite{NEWBOOK}).

\begin{figure}
\includegraphics[scale=0.8]{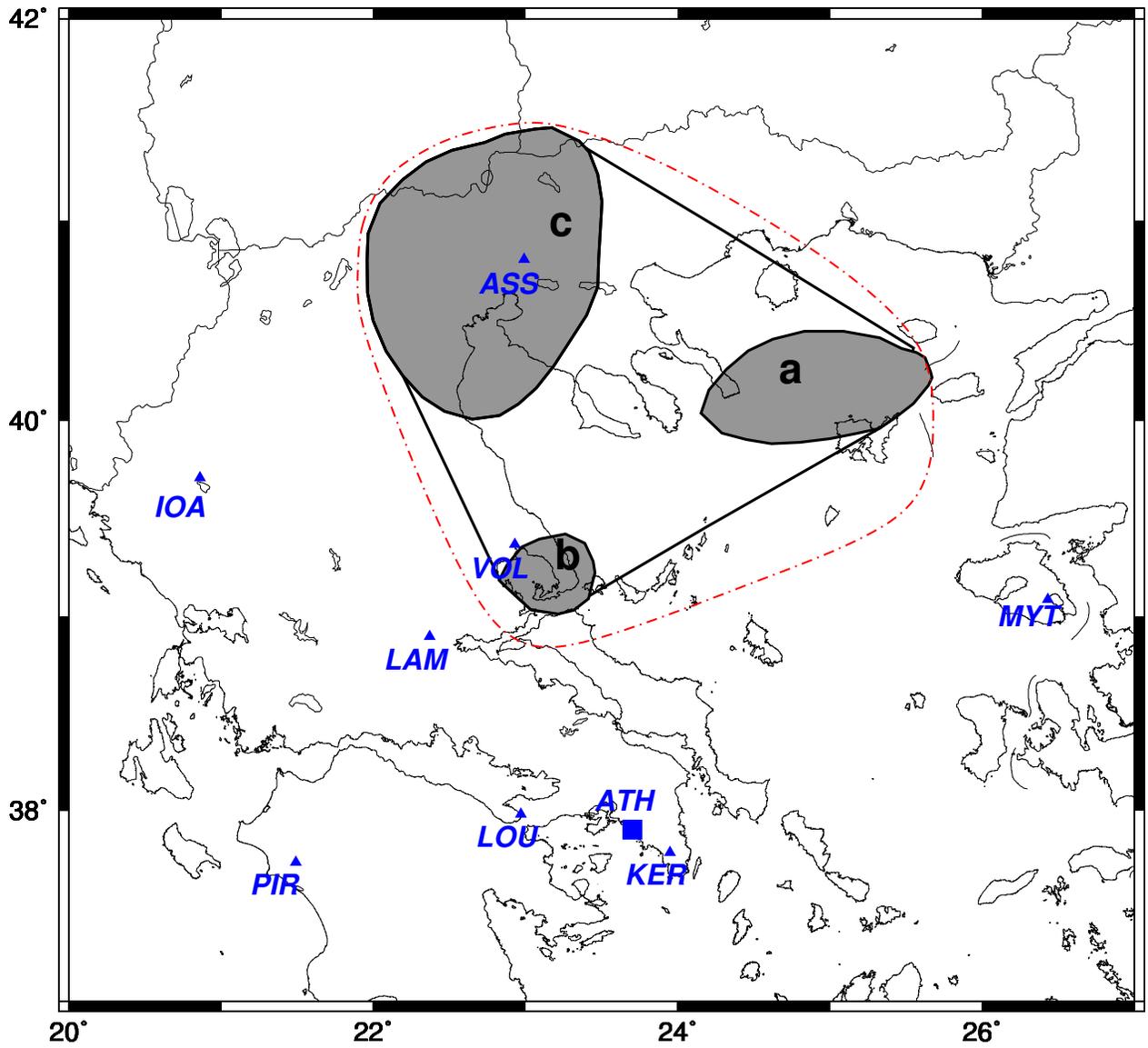}
\caption{Map showing the operating VAN stations (blue triangles). The ASS selectivity map is bounded by the red dashed-dotted line as it was obtained in N. Sarlis, 
{\em Proc. Jpn. Acad. Ser. B} {\bf 89}, 438-445 (2013).} \label{fig7}
\end{figure}

\begin{figure}
\includegraphics[scale=0.7,angle=270]{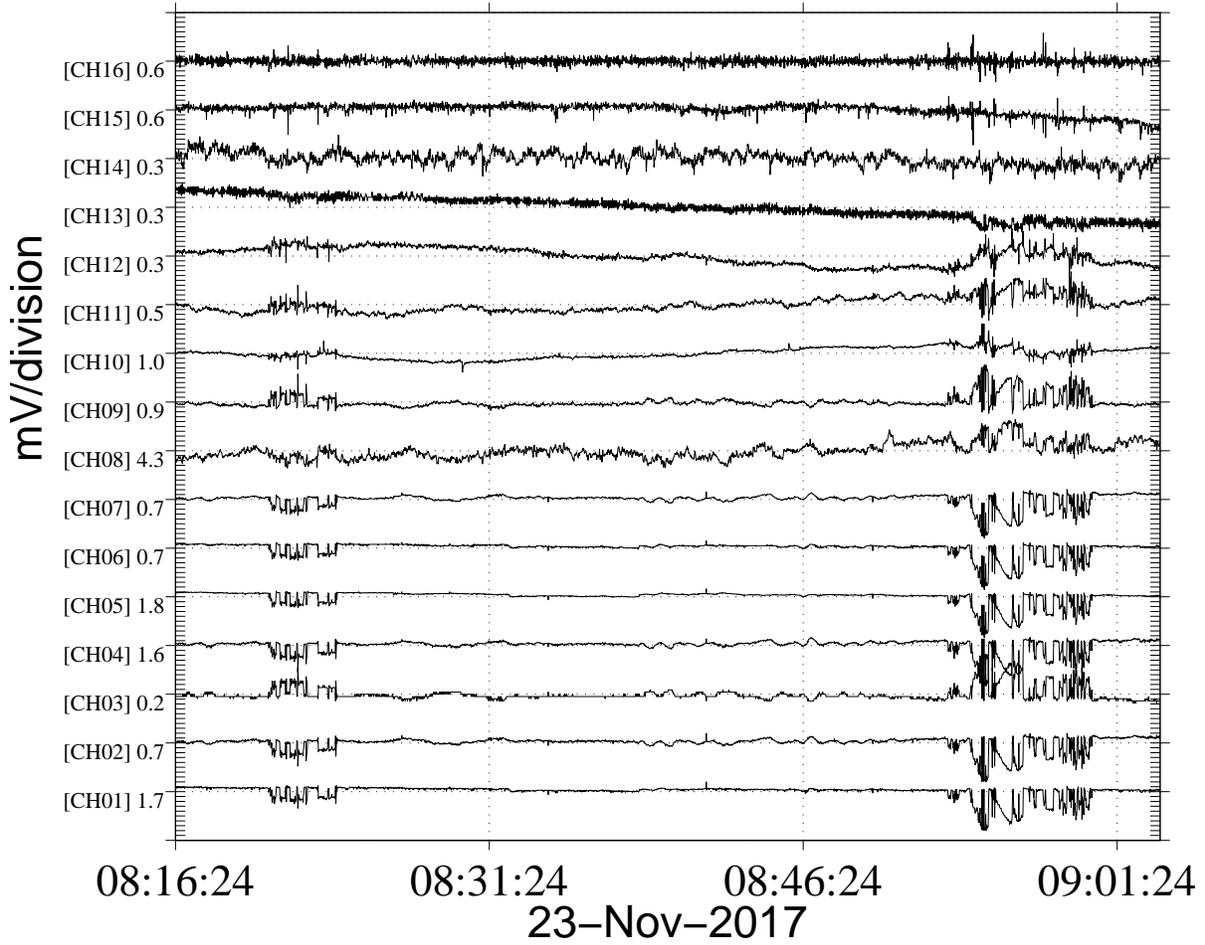}
\caption{The SES activity recorded at ASS station on 23 November 2017. The upper three channels (i.e., CH14, CH15 and CH16) correspond to magnetic field changes recorded by coil magnetometers while the channels labeled CH01 to CH13 to electric field changes  (see the text). } \label{fig8}
\end{figure}

{\it Note added on December 8, 2017:} An SES activity with properties different from those of the
previous SES activity at ASS reported in Ref.\cite{note} was recorded at ASS
station on 23 November 2017 and is depicted in
Fig.\ref{fig8}. The upper 3 channels correspond to the three
components of the magnetic field measured by DANSK coil magnetometers and
the other 13 channels to electric field variations measured by several
short- and long-dipoles of length lying between 100 m and $\approx$9.6 km.
The subsequent seismicity in the ASS selectivity map shown by the red dashed
dotted line in Fig.\ref{fig7} is currently analyzed in natural time in order
to identify the occurrence time of the impending mainshock.

\begin{figure}
\includegraphics[scale=1.0,angle=270]{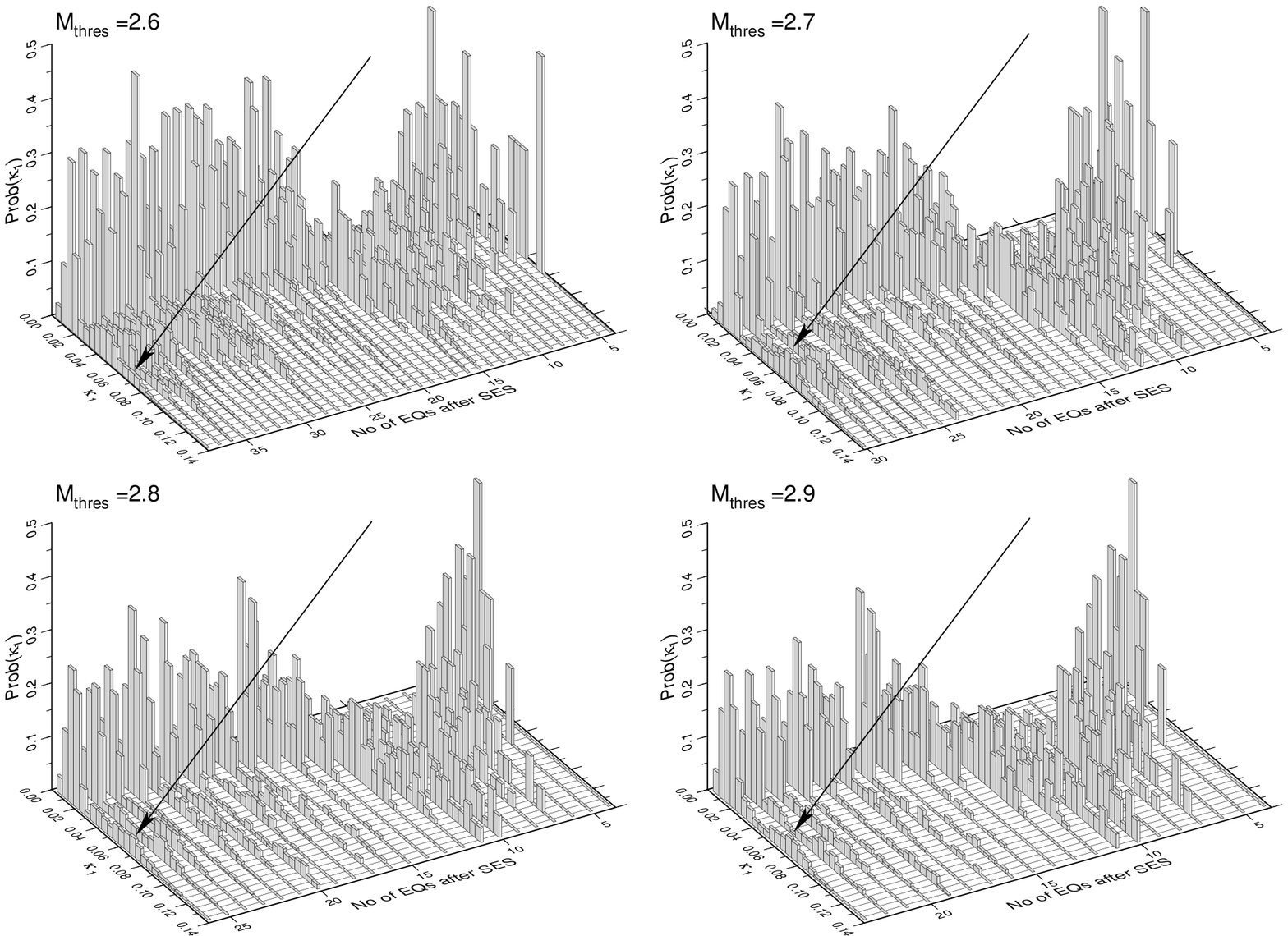}
\caption{Prob($\kappa_1$) versus $\kappa_1$ of the
seismicity(subsequent to the SES activity recorded
at ASS station  depicted in Fig.\ref{fig8})  for
$M_{thres}$=2.6, 2.7, 2.8 and 2.9, within the SES selectivity map of ASS station shown in Fig.\ref{fig7}. The arrows mark the mode of Prob($\kappa_1$) that maximizes at $\kappa_1=0.07$ 
that occurred on 13 February 2018.} \label{fig9}
\end{figure}

{\it Note added on February 14, 2018:} Actually on 2 January 2018 an earthquake of magnitude ML(ATH)4.7, i.e., Ms(ATH)5.2, occurred with an epicenter at 41.2$^o$N, 22.9$^o$E lying within the shaded  region ``c'' of the SES selectivity map of ASS, shown in Fig.\ref{fig7}, at a distance of a few tens of km from the measuring station. Subsequently, the continuation of the natural time analysis of seismicity after the SES activity (Fig.\ref{fig8}) recorded at ASS revealed that on 13 February 2018 upon the occurrence of a number of seismic events  within the shaded region ``a'' as well as 
in the vicinity of the shaded region ``b'' the PDF Prob($\kappa_1$) versus $\kappa_1$ shows a feature (in a similar fashion as Fig.7.23 of Ref.\cite{SPRINGER}) one mode of which maximizes at $\kappa_1=0.070$ exhibiting magnitude threshold invariance (cf. four such examples are depicted for $M_{thres}$=2.6, 2.7, 2.8 and 2.9 in Fig.\ref{fig9}).   

\begin{figure}
\includegraphics[scale=0.75,angle=0]{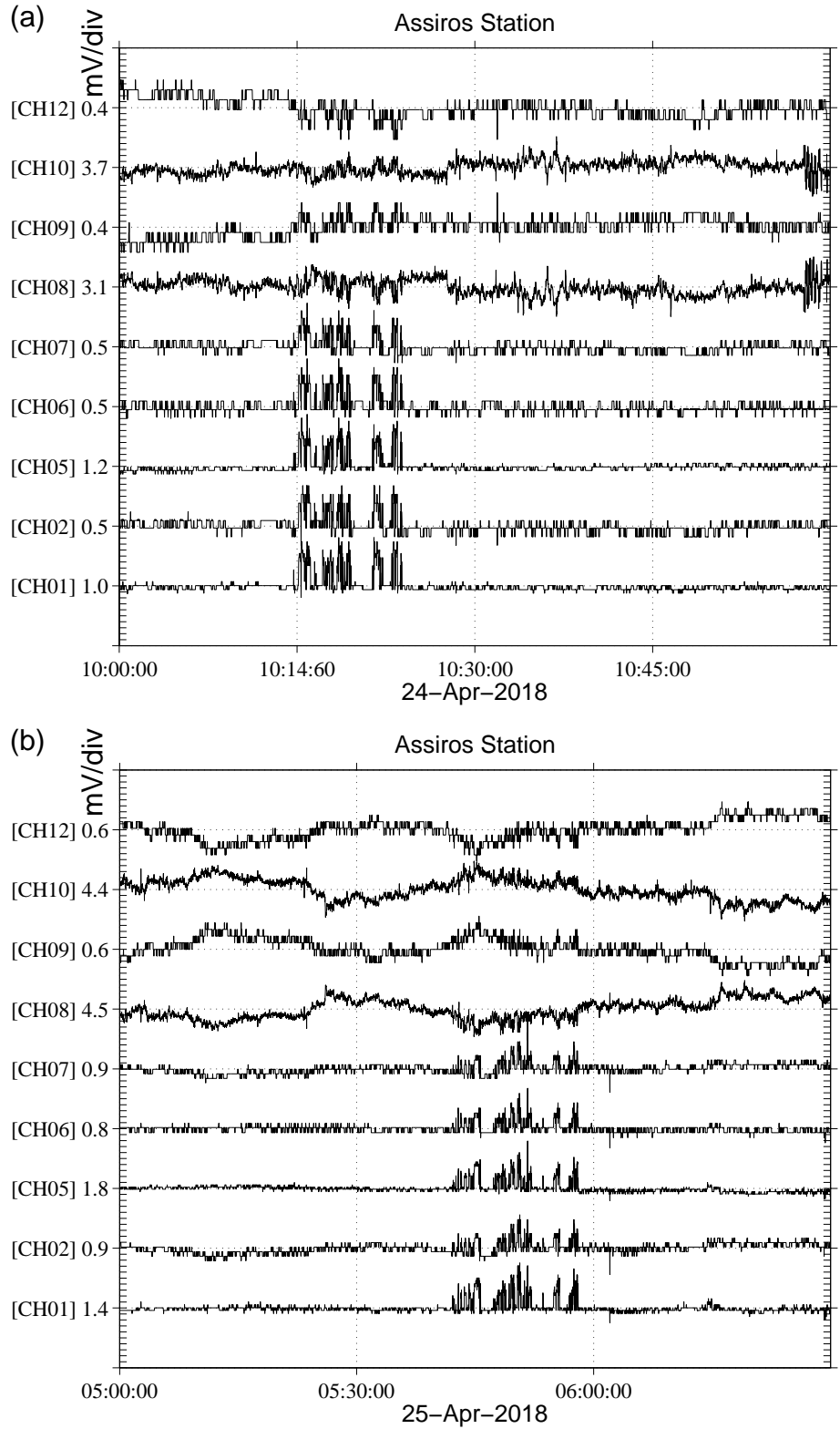}
\caption{The two SES activities recorded at ASS station on 24 April 2018 (a) and 25 April 2018 (b). The ASS selectivity map is shown in Fig.\ref{fig7}. } \label{fig10}
\end{figure}

{\it Note added on June  26, 2018:} On 24 April 2018 and 25 April 2018 two SES activites have been recorded again at ASS (see Fig.\ref{fig10}(a),(b)) but with opposite polarity than
the SES activity in Fig.\ref{fig8}. We recall that an SES activity initiates when a minimum $\beta_{\rm min}$ of the order parameter of seismicity is observed and in addition it has been found  [N. Sarlis et al. {\em Proc.  Natl. Acad. Sci. U.S.A.} {\bf 110} (2013), 13734-13738] that all major shallow earthquakes in Japan were preceded by minima $\beta_{\rm min}$ with lead times around two to three months.

%\bibliography{superef2}
%\bibliographystyle{apsrev}

\end{document}